\documentstyle[12pt,axodraw,epsf]{article}
\begin{document}
\baselineskip 0.6cm
\newcommand{\gsim}{ \mathop{}_{\textstyle \sim}^{\textstyle >} }
\newcommand{\lsim}{ \mathop{}_{\textstyle \sim}^{\textstyle <} }
\newcommand{\vev}[1]{ \left\langle {#1} \right\rangle }
\newcommand{\bra}[1]{ \langle {#1} | }
\newcommand{\ket}[1]{ | {#1} \rangle }
\newcommand{\EV}{ {\rm eV} }
\newcommand{\KEV}{ {\rm keV} }
\newcommand{\MEV}{ {\rm MeV} }
\newcommand{\GEV}{ {\rm GeV} }
\newcommand{\TEV}{ {\rm TeV} }


\begin{titlepage}

\begin{flushright}
CERN-TH/2002-100\\
UT-02-24
\end{flushright}

\vskip 2cm
\begin{center}
{\large \bf  Geometric Origin of Large Lepton Mixing \\
in a Higher Dimensional Spacetime}

\vskip 1.2cm
T.~Watari$^{a,b}$ and T.~Yanagida$^{a,b,}$\footnote{
On leave from University of Tokyo, Japan}

\vskip 0.4cm
$^{a}$ {\it Theory Division, CERN, CH-1211 Geneva 23, Switzerland} \\
$^{b}$ {\it Department of Physics, University of Tokyo, \\
          Tokyo 113-0033, Japan}\\

\vskip 1.5cm
\abstract{
The large mixing in the lepton sector observed in the recent 
neutrino-oscillation experiments strongly suggests that 
nature of left-handed lepton doublets is very different from 
that of left-handed quark doublets. 
This means that there is a big disparity between the matter multiplets 
{\bf 5$^*$}'s and {\bf 10}'s in the SU(5) unified theory. 
We show that this big difference can be explained in a six-dimensional 
spacetime compactified on the {\bf T$^2$}/{\bf Z$_3$} orbifold. 
That is, we propose to put three families of {\bf 5$^*$}'s 
on three equivalent fixed points of the orbifold and 
the three {\bf 10}'s in the two-dimensional bulk. 
We construct an explicit model realizing this situation 
and show that the democratic mass structure in the lepton sector 
is naturally obtained and hence the model explains the observed 
bi-large lepton mixing and simultaneously the required small mixing
$U_{e3}$. 
The mass matrices and mixing  
in the quark sector are also briefly discussed.
}

\end{center}
\end{titlepage}


\section{Introduction}

Bi-large mixing in the lepton sector \cite{Lmix,BP} is one of the most
remarkable features in the standard model.
Contrary to the quark sector, where the mass-diagonalization
matrix of the left-handed up-type quarks almost coincides with that of 
the left-handed down-type quarks,
there is a large discrepancy between the matrix of the left-handed charged 
leptons and that of the neutrinos. 
What is more embarrassing is the small value of $U_{e3}$ \cite{CHOOZ} 
in the lepton flavour mixing matrix.

Various attempts have been made to get a better understanding 
of such flavour structure, yet we do not have a satisfactory picture.
The Froggatt--Nielsen structure is able to explain the bi-large mixing 
assuming less hierarchy for the lepton doublets \cite{lop-sided,anarchy}, 
but an accidental cancellation is necessary between 
${\cal O}(1)$ coefficients to explain the large mixing angle 
of the solar neutrino oscillation \cite{lop-sided} or the small 
$U_{e3}$ \cite{anarchy}.
The democratic mass matrix for the charged leptons 
\cite{democratic1,democratic2} leads to large-angle rotations 
in the mass diagonalization, 
but this does not necessarily mean a large mixing in the W-boson current, 
unless the mass matrix for the Majorana neutrinos is almost diagonal.

A number of tentative solutions are proposed within the framework of 
four-dimensional field theories, where one tries to understand 
the flavour structure as a result of some symmetries.
In this letter, we propose a framework in a higher-dimensional 
spacetime. Here, one can incorporate dynamical effects 
in the mass textures, which cannot be described by symmetries 
of four-dimensional field theories.

The model is described in terms of SU(5) unified theories with 
four-dimensional ${\cal N}=1$ supersymmetry. 
We start from a six-dimensional SU(5) field theory with a 
minimal supersymmetry. A hypermultiplet in the SU(5)-{\bf adj.} 
representation is introduced to cancel the box anomaly arising from the
SU(5) vector multiplet.
Three sets of an anomaly-free combination\footnote{For detail, see the
text.} of hypermultiplets, 
SU(5)-({\bf 10(10$^*$)}$\oplus$3$\times${\bf 5(5$^*$)}), are also introduced.
Two extra dimensions are compactified on a ${\bf T}^2/
{\bf Z}_3\vev{\sigma}$ orbifold.
Three ${\cal N}$=1 SU(5)-${\bf 10}$ chiral multiplets survive 
the orbifold projection, and three ${\bf 5}^*$'s (and ${\bf 1}$'s) 
are further introduced at the three fixed points of 
the ${\bf T}^2/{\bf Z}_3\vev{\sigma}$ orbifold for theoretical consistency.
These multiplets are identified with the three families of quarks, 
leptons (and right-handed neutrinos).
It is quite natural to consider that the three ${\bf 5}^*$'s are
equivalent, since the three fixed points are equivalent to 
one another \cite{previous}.

Origins of SU(5)-{\bf 10}'s and SU(5)-{\bf 5$^*$}'s are totally
different in this model.
The {\bf 10}'s propagate in the whole bulk, while the {\bf 5$^*$}'s and 
{\bf 1}'s localize at three fixed points.
Then, each {\bf 10} has the same wave function at the 
three fixed points, and hence the texture of the charged-lepton mass
matrix will be democratic. On the contrary, 
the Dirac Yukawa matrix for the (left- and right-handed) neutrinos 
and mass matrix of the right-handed Majorana
neutrinos would be almost diagonal (hence so is that of the left-handed
Majorana neutrinos), since a pair of  ${\bf 5}^*$ and {\bf 1} 
is isolated from other pairs of ${\bf 5}^*$ and ${\bf
1}$ at different fixed points.
The diagonal nature of the Majorana neutrino mass matrix is naturally
explained, and hence the bi-large mixing and the smallness of the
$U_{e3}$ follows in the lepton flavour mixing matrix \cite{democratic2}.
The flavour structure of the quark sector is also discussed.

\section{The Model}

An SU(5) vector multiplet of the minimal supersymmetry gives rise to 
box anomalies in the six-dimensional spacetime. 
The Green--Schwarz mechanism \cite{GSmechanism} cannot be used to 
cancel these anomalies.
They are cancelled by introducing a 
hypermultiplet in the {\bf adj.} representation of the SU(5) gauge
group.
One can add one hypermultiplet in the {\bf 10(10$^*$)}
representation\footnote{A hypermultiplet in the {\bf 10(10$^*$)}
representation consists of Kaluza--Klein towers of chiral multiplets in
the {\bf 10} and {\bf 10$^*$} representations of the four-dimensional
${\cal N}=1$ supersymmetry. So does that in the {\bf 5(5$^*$)} 
representation.} together with three in the {\bf 5(5$^*$)} 
without introducing irreducible anomalies\footnote{One has to 
introduce two-form field to cancel reducible anomalies 
through the Green--Schwarz mechanism.} \cite{FK}.
We introduce three sets of this anomaly-free combination in addition to 
the hypermultiplet in the {\bf adj.} representation.

Two extra dimensions are compactified. 
The length scale of the compactified manifold is assumed to be
larger than the Planck length, but smaller than the inverse of the 
GUT scale ($\sim 10^{16}$ GeV). 
Since the minimal supersymmetry in the six-dimensional spacetime 
becomes ${\cal N}=2$ supersymmetry in the four-dimensional spacetime, 
we take an orbifold as the compactified manifold to reduce 
the supersymmetry.
We consider a ${\bf T^2}/{\bf Z_3}\vev{\sigma}$ orbifold. 
This orbifold has three equivalent fixed points, and this is 
why we choose this background geometry \cite{previous}.
The {\bf Z}$_3\vev{\sigma}$ action on the two-dimensional torus ${\bf T}^2$ 
(whose coordinates are described by $z\equiv x_4 + i x_5$) is given by 
\begin{equation}
 \sigma : z \mapsto \sigma \cdot z \equiv \omega^{-2} z,
\label{eq:geo-rotation}
\end{equation}
where the $\sigma$ is a generator of the {\bf Z}$_3\vev{\sigma}$ group 
and $\omega\equiv e^{2\pi i/3}$.
Fields transform under this geometrical rotation as follows:
\begin{equation}
\Psi(x,z,\theta,\Theta)\equiv 
    (\Sigma + \Theta^{\alpha}{\cal W}_{\alpha})(x,z,\theta)
\mapsto \omega^{-2} \Psi(x,\sigma \cdot z,\omega \theta, \omega \Theta)
\label{eq:geo-vec}
\end{equation}
for the SU(5) vector multiplet (in terms of four-dimensional ${\cal N}=2$
supersymmetry), or equivalently,
\begin{eqnarray}
  {\cal W}_{\alpha}(x,z,\theta) \mapsto 
     \omega^{-1} {\cal W}_{\alpha}(x,\sigma \cdot z,\omega \theta), \\
 \Sigma(x,z,\theta) \mapsto 
     \omega^{-2}\Sigma(x,\sigma \cdot z,\omega \theta),  
\end{eqnarray}
in terms of four-dimensional ${\cal N}=1$ supersymmetry.
For hypermultiplets, 
\begin{eqnarray}
 \Phi(x,z,\theta) \mapsto \Phi(x,\sigma \cdot z,\omega \theta), \\
 \bar{\Phi}(x,z,\theta) \mapsto \bar{\Phi}(x,\sigma \cdot z,\omega \theta).
\label{eq:geo-hyp} 
\end{eqnarray}
Here, the chiral multiplets of the four-dimensional ${\cal N}=1$
supersymmetry, $\Phi$ and $\bar{\Phi}$, stand for the hypermultiplets 
($\Phi({\bf adj.})$,$\bar{\Phi}({\bf adj.})$),
($\Phi({\bf 10})$,$\bar{\Phi}({\bf 10^*})$), 
($\Phi({\bf 5})$,$\bar{\Phi}({\bf 5^*})$) ,
($\Phi'({\bf 5})$,$\bar{\Phi}'({\bf 5^*})$) and
($\Phi''({\bf 5})$, $\bar{\Phi}''({\bf 5^*})$).

The orbifold projection selects out states that are invariant
under an action of an orbifold group; this group is generated by
a symmetry transformation of the extra-dimensional space (i.e.
{\bf Z}$_3$, in this case, given in eq.(\ref{eq:geo-rotation})) 
accompanied by twisting fields 
by internal discrete symmetries of the action.
To keep the consistency of the resulting theories 
the internal symmetry to be used for the twisting of fields should 
have the same degree as that of the space symmetry (i.e. {\bf Z}$_3$).

The first candidate of such an internal symmetry is the SU(2) R symmetry.
(The SU(2) R symmetry of the minimal supersymmetry of the six-dimensional
spacetime is what would become an SU(2) R symmetry of the ${\cal N}=2$
supersymmetry in the four-dimensional spacetime when one would simply take
a toroidal compactification.)
This SU(2) symmetry does not have any anomaly with gauge groups;
both SU(2)[SU(5)]$^3$ and SU(2)[gravity]$^3$ box anomalies vanish.
Thus, we use a {\bf Z}$_3$ subgroup included 
in the maximal torus (U(1) subgroup) of the SU(2) R symmetry.
The SU(5) vector multiplet transforms under the maximal torus as
\begin{equation}
 \Psi(x,z,\theta,\Theta) \rightarrow
     \Psi(x,\sigma \cdot z,e^{-i\alpha}\theta,e^{i\alpha}\Theta),
     \qquad \alpha \in {\bf R},
\end{equation}
or equivalently,
\begin{eqnarray}
 {\cal W}_{\alpha}(x,z,\theta) \rightarrow 
          e^{i\alpha}{\cal W}_{\alpha}(x,\sigma \cdot z,e^{-i\alpha}\theta),
\label{eq:R-rotation1} \\
 \Sigma(x,z,\theta) \rightarrow
          \Sigma(x,\sigma \cdot z,e^{-i\alpha}\theta) 
\label{eq:R-rotation2}. 
\end{eqnarray} 
At the same time, the hypermultiplets transform as
\begin{eqnarray}
 \Phi(x,z,\theta) \rightarrow
             e^{i\alpha} \Phi(x,\sigma \cdot z,e^{-i\alpha}\theta), 
\label{eq:R-rotation3}\\
 \bar{\Phi}^{\dagger}(x,z,\bar{\theta}) \rightarrow
    e^{-i\alpha}\bar{\Phi}^{\dagger}(x,\sigma \cdot z,e^{i\alpha}\bar{\theta})
\label{eq:R-rotation4}. 
\end{eqnarray}
Since the {\bf Z}$_3$ space rotation changes the Grassmann coordinate 
$\theta$ into $\omega \theta$ (eqs. (\ref{eq:geo-vec}--\ref{eq:geo-hyp})), 
the {\bf Z}$_3\vev{\sigma}$ transformation keeps $\theta$ invariant 
if the space rotation is accompanied by a twisting under the SU(2) R symmetry 
with $e^{-i\alpha} =\omega^{-1}$. 
Then, the orbifold projection conditions do not make any 
discrimination between bosonic and fermionic components in the same ${\cal
N}=1$ multiplets, and hence the ${\cal N}=1$ supersymmetry is kept unbroken.
For the vector multiplet, the orbifold projection conditions are 
now given by
\begin{eqnarray}
 {\cal W}_{\alpha}(x,z,\theta) = 
     \left( \sigma :  {\cal W}_{\alpha}(x,z,\theta) \mapsto 
                      {\cal W}_{\alpha}(x,\sigma \cdot z,\theta) \right), \\
 \Sigma(x,z,\theta) = 
     \left( \sigma :  \Sigma(x,z,\theta) \mapsto 
                     \omega^{-2} \Sigma(x,\sigma \cdot z,\theta) \right).
\end{eqnarray}

The SU(2) R symmetry, however, is not sufficient to yield a
phenomenologically interesting model.
This is because all the hypermultiplets would be, then, under the conditions
\begin{eqnarray}
 \Phi(x,z,\theta) = \omega \Phi(x,\sigma \cdot z,\theta), \\
 \bar{\Phi}^{\dagger}(x,z,\bar{\theta}) =
    \omega^{-1}\bar{\Phi}^{\dagger}(x,\sigma \cdot z,\bar{\theta}),  
\end{eqnarray}
and hence no Kaluza--Klein zero mode would survive.
We therefore introduce a twisting by an additional global U(1) symmetry 
in the orbifold projection conditions.
This U(1) symmetry should not be violated even quantum mechanically,
and hence the anomaly with gauge fields should vanish.
U(1)[gravity]$^3$ always vanishes, and U(1)[SU(5)]$^3$ 
vanishes if one takes
the charge assignment of ``fiveness''\footnote{``Fiveness'' U(1)
symmetry can be gauged in the bulk. Reducible anomalies can be cancelled
by the Green--Schwarz mechanism.} given in Table \ref{tab:fiveness}.
(A suitable linear combination of this U(1)$_{\rm fiveness}$ and the
U(1)$_{\rm Y}$ of the standard model is nothing but the U(1)$_{\rm B-L}$.)
The orbifold projection conditions of the hypermultiplets are taken as
\begin{eqnarray}
 \Phi(x,z,\theta) = 
     \left( \sigma :  \Phi(x,z,\theta) \mapsto 
             \omega^q \omega \Phi(x,\sigma \cdot z,\theta) \right), 
\label{eq:final-hyper-transf}\\
 \bar{\Phi}^{\dagger}(x,z,\bar{\theta}) = 
   \left( \sigma : \bar{\Phi}^{\dagger}(x,z,\bar{\theta}) \mapsto 
    \omega^q \omega^{-1}\bar{\Phi}^{\dagger}(x,\sigma \cdot z,\bar{\theta}) 
   \right),
\end{eqnarray}
where $q$ is the fiveness charge. 

Kaluza--Klein zero modes that survive this orbifold projection conditions 
are summarized in terms of four-dimensional ${\cal N}=1$ 
supersymmetry as follows:
an SU(5) vector multiplet and three sets of chiral multiplets\footnote{
We also use the same nomenclature $\Phi({\rm repr.})$ for the Kaluza--Klein
zero modes, which has been used so far as six-dimensional fields 
that depend on the coordinates $z$.} 
$\Phi({\bf 10})$, $\Phi'({\bf 5})$ and $\Phi''({\bf 5})$.
Then, the [SU(5)]$^3$ triangle anomaly arises at all three fixed points, 
but this anomaly is cancelled by introducing  
three chiral multiplets in the {\bf 5}$^*$ representation 
at each fixed point\footnote{
One might expect that a more fundamental theory 
would provide these particles (e.g. as twisted sector fields in
string theories).}.
We denote these three chiral multiplets at a fixed point 
as $X({\bf 5}^*)$, $X'({\bf 5}^*)$ and $X''({\bf 5}^*)$.
The fiveness charge is assigned to these fields\footnote{The reason of 
this assignment for the $X'$({\bf 5}$^*$) and $X''$({\bf 5}$^*$) will be clear 
in the next paragraph. The charge of X({\bf 5}$^*$) is determined so
that the triangle anomaly U(1)$_{\rm fiveness}$[SU(5)]$^2$ vanishes at
each fixed point.} as 3 for $X$({\bf 5}$^*$)  
and $-2$ for $X'$({\bf 5}$^*$) and $X''$({\bf 5}$^*$).

The three sets of the Kaluza--Klein zero modes $\Phi'({\bf 5})$
and $\Phi''({\bf 5})$ form vector-like pairs with the $X'({\bf 5}^*)$
and $X''({\bf 5}^*)$ at all three fixed points,
and hence we expect that their mass terms would be generated.
The chiral matter content consists of three Kaluza-Klein zero modes 
$\Phi({\bf 10})$ from the bulk and three $X({\bf 5}^*)$'s 
 from three fixed points (one $X({\bf 5}^*)$ from one fixed point).
We identify these particles with the three families of quarks and leptons
in the SU(5)-unified theories.
The observed neutrino masses 
suggest right-handed neutrinos below the GUT scale \cite{see-saw}. 
Thus, we also introduce the three families of right-handed neutrinos 
at the three fixed points\footnote{When the ``fiveness'' U(1) symmetry
is gauged in the bulk, then, the triangle anomaly cancellation 
of the ``fiveness'' symmetry at orbifold fixed points 
also requires the right-handed neutrinos, i.e. right-handed neutrinos are
also required at fixed points from theoretical consistencies.}. 
They are denoted as $X({\bf 1})$, and we expect that they have the
same origin and hence much the same transformation property 
as the $X({\bf 5^*})$. 
The fiveness charge\footnote{We should note 
here that the triangle anomaly U(1)$_{\rm fiveness}$[gravity]$^2$ 
is cancelled out by this introduction of the right-handed neutrino 
at each fixed point.} of this $X({\bf 1})$ is $-5$.
The U(1) fiveness should be 
broken below the GUT scale so that the right-handed neutrinos 
acquire masses.

There should be Higgs particles that give masses to quarks and leptons.
We also need a sector that breaks the SU(5) symmetry down to the
standard-model gauge group.
In order to keep the geometrical equivalence of the three fixed points,
we assume that the Higgs particles (H({\bf 5}) and $\bar{H}({\bf 5}^*)$) 
and the SU(5)-breaking sector is localized just at the centre of the three
fixed points (see Fig. \ref{fig:middle}).
Such localized sectors in the bulk space
should appear without any theoretical inconsistencies.
Brane solutions of the supergravity give such examples of dynamical
localization of supersymmetric gauge theories\footnote{One might then
consider that there would be unwanted Nambu--Goldstone modes
corresponding to the branes' motion in the smooth space. 
However, since the {\em continuous} translational symmetry 
has already been lost in the orbifold geometry, 
it can be expected that this violation of the symmetry 
gives masses to the Nambu--Goldstone modes 
through non-perturbative effects.
Discussion on such non-perturbative dynamics, however, 
is beyond the scope of this paper.}, and hence we refer 
to these sector as ``centre brane''.
An SU(5)-breaking model can be found in \cite{ss-IY}, where
coloured Higgs particles are given masses of the order of the GUT scale 
naturally, while Higgs doublets remain massless.
The brane-world realization of this model is also discussed 
in \cite{ss-IWY,ss-WY}.   
The four-dimensional ${\cal N}=2$ supersymmetric multiplet structure 
of this model \cite{ss-HY,ss-IWY} might be of relevance in such 
a realization. 

Finally, we show that this extra-dimensional model is able to provide
a suitable R symmetry. 
This fact provides quite a non-trivial consistency check for 
the phenomenological model building in the extra-dimensions.
Moreover, an R symmetry (mod 4) is indispensable
to the above SU(5)-breaking model \cite{ss-IY,KMY}. 

The maximal torus of the SU(2) R symmetry can be
preserved in the theory on the orbifold, since it commutes with the
orbifold group. 
Since it rotates the Grassmann coordinates $\theta$ as in
eqs. (\ref{eq:R-rotation1}--\ref{eq:R-rotation4}), it is an R symmetry
of the four-dimensional ${\cal N}=1$ supersymmetry.
The three chiral multiplets $\Phi({\bf 10})$'s transform as
in eq. (\ref{eq:R-rotation3}) and hence it carries R charge 1, as desired
phenomenologically.
At the same time, $\Phi'({\bf 5})$ and $\Phi''({\bf 5})$ also carry 
R charge 1.
Then, the chiral multiplets $X'({\bf 5^*})$ and $X''({\bf 5^*})$ at
the fixed points are required to have R charge 1 so that the mass terms
\begin{equation}
 W \propto \Phi^{'('')}({\bf 5}) X^{'('')}({\bf 5^*}) 
\end{equation}
are allowed by the R symmetry.
Now we assume that the remaining multiplets at the fixed points, 
namely $X({\bf 5^*})$ and $X({\bf 1})$, 
also have the same R charge as the fixed-point fields $X'({\bf 5}^*)$
and $X''({\bf 5}^*)$.  
Then, this means that quarks and leptons in SU(5)-{\bf 5}$^*$ and 
right-handed neutrinos have R charge 1, 
which is again the desired assignment \cite{ss-IY,ss-IWY,ss-WY}. 
As for the Higgs multiplets and fields in the SU(5)-breaking sector,
there is no ``top down'' way to determine their transformation 
property under the maximal torus of the SU(2) R symmetry.
Thus, we simply expect that their charge assignment is suitably
realized. 
We assume that the maximal torus of the SU(2) R symmetry is
broken down to the mod 4 R symmetry as required in \cite{ss-IY,KMY}.

\section{Mass Matrices}

We consider that all operators are generated non-perturbatively, unless 
they are forbidden by symmetries.
We assume that such symmetries are ${\cal N}=1$ supersymmetry,
the (mod 4)-R symmetry and flavour symmetries, 
which we discuss below.

The flavour symmetries for the fixed-point fields and for the fields in the
bulk are independent, since their origins are different.  
That is, ($X({\bf 5}^*)$'s and $X({\bf 1})$'s) and $\Phi({\bf 10})$'s have
independent flavour structures. 
We denote the family indices as $X({\bf 5}^*)_i$, 
$X({\bf 1})_i$ and $\Phi({\bf 10})_a$, where $i= 1,2,3$ correspond 
to three fixed points and $a=1,2,3$ to three hypermultiplets 
in the bulk.

First of all, we discuss the flavour symmetry of the fields on the fixed
points, i.e. $X({\bf 5}^*)_i$'s and $X({\bf 1})_i$'s. 
Flavour symmetry for the $\Phi({\bf 10})_a$'s is briefly 
discussed later. 

One can see in the Fig. \ref{fig:middle} that 
the {\bf T}$^2$/{\bf Z}$_3\vev{\sigma}$ orbifold possesses 
{\bf Z}$_3\vev{\tau}$ translational symmetry generated by $\tau$. 
The translation results in a cyclic permutation 
of the three fixed points, 
under which fixed-point fields transform as
\begin{eqnarray}
 \tau :& X({\bf 5}^*)_i &\mapsto X({\bf 5}^*)_{i+1}, \\
       & X({\bf 1})_i   &\mapsto X({\bf 1})_{i+1},
\end{eqnarray}
where $i=4$ is identical to $i=1$. 
Since the $\tau$ is nothing but a translation, the bulk Kaluza--Klein 
zero modes $\Phi({\bf 10})_a$ are invariant under the {\bf Z}$_3\vev{\tau}$
transformation.
 
Now let us discuss the neutrino mass textures.
The {\bf Z}$_3\vev{\tau}$ translational symmetry allows the
following degrees of freedom in the Dirac Yukawa couplings  
and Majorana mass terms of neutrinos, respectively:
\begin{eqnarray}
W= y_{ij}^D X({\bf 5}^*)_i X({\bf 1})_j H({\bf 5})  
        \qquad \qquad y_{ij}^D = c^D \left( 
               \begin{array}{ccc}
                1 & a & b \\ b & 1 & a \\ a & b & 1
               \end{array}\right),  \label{eq:Dirac}\\
W = h_{ij}^{R} M_R X({\bf 1})_i X({\bf 1})_j 
         \qquad \qquad h_{ij}^{R} = c^{R} \left(
                \begin{array}{ccc}
                 1 & a' & a' \\ a' & 1 & a' \\ a' & a' & 1
                \end{array}\right), \label{eq:R-Majorana}
\end{eqnarray}
where $M_R$ is of the order of the mass scale of right-handed neutrinos,
and the Majorana mass terms of the left-handed neutrinos is obtained 
through the see-saw mechanism \cite{see-saw} as 
\begin{eqnarray}
 W = \frac{1}{M_R}h_{il}^{L} (X({\bf 5^*})_i H({\bf 5})) 
(X({\bf 5^*})_l H({\bf 5})),  \\
h_{il}^{L} = y^D_{ij} (h^{R})^{-1}_{jk} y^D_{lk} 
= (c^D)^2 (c^{R})^{-1} 
  \left(\begin{array}{ccc}
     1 & \kappa & \kappa \\ \kappa & 1 & \kappa \\ \kappa & \kappa & 1
        \end{array}\right). 
\end{eqnarray}
Off-diagonal elements are relatively suppressed as 
$a,b,a' \sim e^{-M_* l} \ll 1$, provided $M_* l \gg 1$,
where the $M_*$ is the fundamental scale of the theory and $l$ is the
typical length scale of the orbifold geometry \cite{AHD}.
This is because three $X({\bf 5^*})_i$'s and $X({\bf 1})_i$'s 
are localized at a fixed point distant from others.
As a result, the Majorana mass matrix of the left-handed neutrinos   
is also almost diagonal ($\kappa \ll 1$).
Notice that the higher-dimensional configuration ($M_* l \gg 1$) suppresses 
the off-diagonal terms, 
which cannot be forbidden by the $S_3$ (or ${\bf Z}_3$) symmetry 
in the democratic ansatz in the four-dimensional spacetime \cite{democratic2}.
This is the most crucial point in this paper.
Although there are preceding trials to interpret various properties of
mass texture in terms of geometry \cite{AS}, it should be emphasized
here that the localization of $X({\bf 5}^*)_i$'s and $X({\bf 1})_i$'s 
at suitable positions is not a choice by hand but rather an inevitable
consequence of theoretical consistencies. 

The left-handed Majorana neutrinos have almost diagonal mass matrix, even
when the ${\bf Z}_3\vev{\tau}$ symmetry is slightly broken; 
the breaking effects also have the extra $e^{-M_* l}$
suppression in off-diagonal matrix elements.
Since the mass matrix of the charged leptons is subject to 
the ${\bf Z}_3\vev{\tau}$ symmetry (with small breaking effects 
of this symmetry), large angle rotation between charged leptons 
is necessary for the mass diagonalization \cite{democratic2}. 
Thus, the large mixing follows, in general, without specifying how 
the ${\bf Z}_3\vev{\tau}$ symmetry is broken or without assuming 
any flavour structure in the $\Phi({\bf 10})_a$'s.

Before proceeding further, we briefly comment 
on the flavour symmetry of the $\Phi({\bf 10})_a$'s.
Since the bulk Lagrangian\footnote{It is impossible to write down the whole 
Lagrangian of the theory when the six-dimensional field theory 
with the minimal supersymmetry has more than one tensor multiplets 
\cite{tensor}.
In that case, what is discussed here means that the field equations 
that result from this ``Lagrangian'' possess an accidental SU(3)
symmetry.} 
is restricted by the higher-dimensional Lorentz symmetry and an 
extended supersymmetry, the leading interaction for the $\Phi({\bf 10})_a$'s 
is \cite{Arkani}
\begin{equation}
 W = 2  \bar{\Phi}({\bf 10^*})^a(z) 
        \left( \bar{\partial_z} - \frac{g}{\sqrt{{2}}}\Sigma(z)\right) 
      \Phi({\bf 10})_a(z),
\end{equation}
where $g$ is the SU(5)-gauge coupling constant.
This interaction is flavour-universal and hence an SU(3) accidental 
symmetry exists.
The $\Phi({\bf 10})_a$'s form a {\bf 3} representation of the SU(3) symmetry. 
This SU(3) flavour symmetry should be broken so that the Yukawa
couplings are allowed.
When its breaking is encoded\footnote{The symmetry and its
breaking pattern given here is just an example. Most of our discussion 
still holds even if the SU(3) flavour symmetry is broken in a different
way. We take this example just as an illustration.} by a spurion field 
$v^a$ in a {\bf 3}$^*$ representation of the SU(3) 
with all three components of order 1, the superpotential \cite{TWY}
\begin{equation}
 W = c  X({\bf 5^*})_i v^a \Phi({\bf 10})_a  \bar{H}({\bf 5^*}), \qquad 
W = c' v^a \Phi({\bf 10})_a v^b \Phi({\bf 10})_b H({\bf 5}),
\end{equation}
leads to rank 1 mass matrices of the democratic type.
The mixing angles of the CKM matrix vanish, since the
mass-diagonalization matrices of the $\Phi({\bf 10})_a$'s are exactly 
the same in the up-type Yukawa coupling and 
in the down-type Yukawa coupling in the absence of breakings of the
democratic form \cite{democratic1}\footnote{The mixing angle between the 
first and the second families does not make sense without the breakings,
since the quarks of these two families are massless.}.

Finally, let us discuss the breaking effects to these flavour symmetry.
First of all, let us assume that the centre brane is displaced slightly 
toward the fixed point 3 (see Fig. \ref{fig:middle}). 
Then, the Majorana mass matrix of the left-handed neutrinos becomes
non-degenerate as
\begin{equation}
h_{il}^{L} = y^D_{ij} (h^{R})^{-1}_{jk} y^D_{lk} 
= (c^D)^2 (c^{R})^{-1} 
  \left(\begin{array}{ccc}
     1 & \kappa & \kappa \\ \kappa & 1 & \kappa \\ \kappa & \kappa & 1
        \end{array}\right) \longrightarrow
 (c^D)^2 (c^{R})^{-1} 
  \left(\begin{array}{ccc}
     1 & \kappa & \kappa' \\ \kappa & 1 & \kappa' \\
      \kappa' & \kappa' & 1+\delta,
        \end{array}\right),
\end{equation}
where $(\kappa'-\kappa) \sim \delta \kappa$ as discussed before, 
and we assume that $\kappa \ll \delta \lsim 1$. Further breaking ($\gsim 
\kappa $) will resolve the remaining degeneracy, leading to 
the $\Delta m^2$ of the solar neutrino oscillation small compared with 
the $\Delta m^2$ of the atmospheric neutrino oscillation.
Secondly, we assume, for example\footnote{This particular example is
taken just for an illustration of our idea.}, 
that the $\Phi({\bf 10})_3$ is the most sensitive 
to the displacement of the centre brane among all the $\Phi({\bf 10})_a$'s.
Then, the charged-lepton mass matrix becomes
\begin{equation}
c v_a = c \left( \begin{array}{ccc}
	  v_1 & v_2 & v_3 \\  v_1 & v_2 & v_3 \\  v_1 & v_2 & v_3 \\ 
	       \end{array}\right) \longrightarrow
        c \left( \begin{array}{ccc}
	 v_1 & v_2 & v_3 (1+\epsilon) \\v_1 & v_2 & v_3 (1+\epsilon) \\
         v_1 (1+\delta') & v_2 (1+\delta') & v_3 (1+\delta'+\epsilon')
	       \end{array}\right),
\label{eq:Z2texture}
\end{equation} 
where $\delta^{'} \sim \delta$ and $\epsilon \sim \epsilon'$,
and the muon acquires a mass suppressed by $\sim \epsilon \delta$
relative to the mass of the tau lepton. 
If the breaking dynamics 
still preserves the equality between $X({\bf 5}^*)_1$ and $X({\bf
5}^*)_2$ as in eq. (\ref{eq:Z2texture}), then the small $U_{e3}$ is also 
obtained \cite{democratic2}. At least, the displacement of the centre brane
to the fixed point 3 does not make any difference in the distances
between the Higgs particle (centre brane) and fields at fixed points, 
$X({\bf 5}^*)_1$ and $X({\bf 5}^*)_2$. We assume that this equality
holds\footnote{Even if there is a difference between the matrix elements 
of the $X({\bf 5}^*)_1$'s row and the $X({\bf 5}^*)_2$'s row, some of
them can be absorbed by rescaling and rephasing of the $X({\bf 5}^*)_1$
and $X({\bf 5}^*)_2$, and it is possible that the difference
that cannot be absorbed is of order $\delta^2$. 
Then, the $U_{e3}$ is still sufficiently small.}.

The mass matrix for the up-type quarks may now be written as
\begin{equation}
c' v_a v_b = c' \left(\begin{array}{ccc}
	   v_1 v_1 & v_1 v_2 & v_1 v_3 \\
	   v_2 v_1 & v_2 v_2 & v_2 v_3 \\
	   v_3 v_1 & v_3 v_2 & v_3 v_3 
		\end{array}\right) \rightarrow
c' \left(\begin{array}{ccc}
	   v_1 v_1 & v_1 v_2 & v_1 v_3 (1+\epsilon'')\\
	   v_2 v_1 & v_2 v_2 & v_2 v_3 (1+\epsilon'')\\
	   v_3 v_1 (1+\epsilon'') & v_3 v_2 (1+\epsilon'') 
                  & v_3 v_3 (1+\epsilon^{'''}) 
		\end{array}\right) ,
\label{eq:1010texture}
\end{equation}
where $\epsilon'',\epsilon^{'''} \sim \epsilon$ and $m_c / m_t 
\sim \epsilon^2$, and that for the down-type quarks is given by 
eq. (\ref{eq:Z2texture}).
Mass hierarchy and mixing in the quark sector are obtained in a similar
way to \cite{democratic1,democratic2}. It is clear from 
(\ref{eq:Z2texture}) and (\ref{eq:1010texture}) that small angles are
derived in the CKM matrix, since the diagonalization matrices for 
the up- and the down-type left-handed quarks almost coincide.

Effective Yukawa coupling of down-type quarks and charged leptons $c$ is
suppresssed by $e^{-M_* l}$ compared with the up-type Yukawa coupling
$c'$. We expect $c \sim 10^{-1} c'$, and hence the tan $\beta$ is not so
large. Off-diagonal elements of the left-handed Majorana neutrino mass
matrix are also suppressed by $\kappa$ relatively to the diagonal elements.
Natural explanation of bi-large mixing, on the other hand, requires 
$\kappa \lsim 10^{-2}$. 
When the off-diagonal elements of Eq. (\ref{eq:Dirac}) and
Eq. (\ref{eq:R-Majorana}) are generated by exchanging two massive
particles, $\kappa$ is given by a ratio $(e^{-M_* l})(e^{-M_*
l})/(e^{-M_* l})$, slight difference between three $M_*$'s in this
ratio easily leads to $\kappa \sim 10^{-2}$ rather than $\kappa \sim
10^{-1}$.
When the off-diagonal elements are generated with suppression factors 
$e^{- M_*^2 ({\rm area})}$, then there is no wonder that the $\kappa$ 
is smaller than the ratio $c/c'$.

\section*{Acknowledgements}
T.W. thanks the Japan Society for the Promotion of Science for
financial support. This work is supported by Grant-in-Aid for Scientific
Research (S) 14102004 (T.Y.).

\newpage
\begin{figure}
\begin{center}
\begin{picture}(300,300)(-150,-150)
%
\Line(-150,78)(150,78)
\Line(-150,0)(150,0)
\Line(-150,-78)(150,-78)
%
\Vertex(-90,52){2} \Vertex(-90,0){2} 
\Vertex(  0,52){2} \Vertex(  0,0){2} \Vertex(  0,-52){2}
                    \Vertex( 90,0){2} \Vertex( 90,-52){2}

\Vertex(-135,78){2}                                        
\Vertex(-45,78){2} \Vertex(-45,26){2} \Vertex(-45,-26){2} \Vertex(-45,-78){2}
\Vertex( 45,78){2} \Vertex( 45,26){2} \Vertex( 45,-26){2} \Vertex( 45,-78){2}
                                                          \Vertex( 135,-78){2}
%
\Line(-158,-39)(-113,-117)
\Line(-158,117)(-23,-117)
\Line(-68,117)(68,-117)
\Line(23,117)(158,-117)
\Line(113,117)(158,39)
%
\Text(-135,78)[lb]{3}  \Text(-45,78)[lb]{3}  \Text( 45,78)[lb]{3}
\Text(-90,52)[lb]{2}  \Text(  0,52)[lb]{2}
\Text(-45,26)[lb]{1}  \Text( 45,26)[lb]{1} 
\Text(-90,0)[lb]{3}   \Text(  0,0)[lb]{3}   \Text( 90,0)[lb]{3}
\Text(-45,-26)[lb]{2} \Text( 45,-26)[lb]{2}
\Text(  0,-52)[lb]{1} \Text( 90,-52)[lb]{1}
\Text(-45,-78)[lb]{3} \Text( 45,-78)[lb]{3} \Text( 135,-78)[lb]{3}
%
\Text(-120,52)[c]{$\circ$}\Text(-30,52)[c]{$\circ$}\Text(60,52)[c]{$\circ$} 
\Text(-75,26)[c]{$\circ$}\Text(15,26)[c]{$\circ$}
\Text(-120,0)[c]{$\circ$}\Text(-30,0)[c]{$\circ$}\Text(60,0)[c]{$\circ$}
\Text(-75,-26)[c]{$\circ$}\Text(15,-26)[c]{$\circ$}\Text(105,-26)[c]{$\circ$}
\Text(-30,-52)[c]{$\circ$}\Text(60,-52)[c]{$\circ$}
\Text(15,-78)[c]{$\circ$}\Text(105,-78)[c]{$\circ$}
\Line(15,26)(10,18) \Line(10,18)(10,22) \Line(10,18)(13,20)
%
\ArrowLine(35,-72)(10,-58)     \ArrowLine(5,-72)(-20,-58)
\ArrowLine(-10,-46)(-35,-32)    \ArrowLine(-40,-46)(-65,-32)
\ArrowLine(-55,-20)(-80,-6)    \ArrowLine(-85,-20)(-110,-6)
\Text(-8,-66)[rt]{$\tau$}
\Text(-53,-40)[rt]{$\tau$}
\Text(-98,-14)[rt]{$\tau$}
%
%
\end{picture}
\end{center}
\caption{A picture of the ${\bf T}^2/{\bf Z}_3\vev{\sigma}$ geometry. 
A unit cell of the ${\bf T}^2$ torus is described by parallel lines.
Three fixed points are described by $\bullet$ labelled 123, on which 
an SU(5)-{\bf 5}$^*$ and a right-handed neutrino are localized.
We assume that the Higgs particles and the SU(5)-breaking sector
are at the $\circ$ in the figure, which we call the ``centre
 brane''. Although there are three $\circ$'s 
within the unit cell of the ${\bf T}^2$, there is only one in the
 fundamental domain of the {\bf T}$^2$/{\bf Z}$_3\vev{\sigma}$ orbifold.
The ${\bf Z}_3\vev{\tau}$ translational symmetry that leads to the
 flavour symmetry of the fields on fixed points is described by straight
arrow lines. 
A tiny arrow on the centre brane shows its displacement toward the fixed 
 point 3 that leads to the {\bf Z}$_3\vev{\tau}$-symmetry breaking. 
}
\label{fig:middle}
\end{figure}
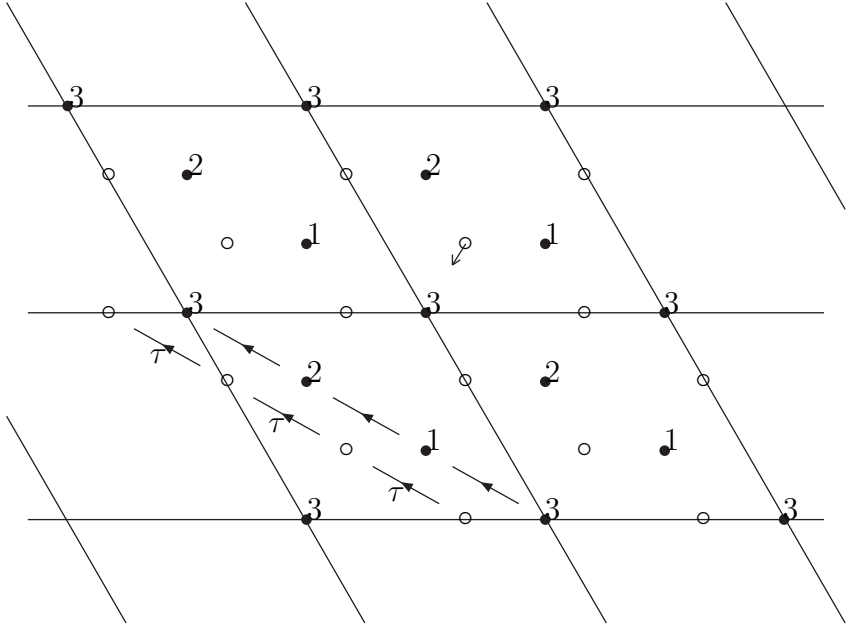
\begin{table}
 \begin{tabular}{c|ccccc}
  Fields & $\Phi({\bf adj.})$,$\bar{\Phi}({\bf adj.})^{\dagger}$ &
                    $\Phi({\bf 10})$,$\bar{\Phi}({\bf 10^*})^{\dagger}$ &
                   $\Phi({\bf 5})$,$\bar{\Phi}({\bf 5^*})^{\dagger}$&
        $\Phi'({\bf 5})$,$\bar{\Phi}'({\bf 5^*})^{\dagger}$  &
         $\Phi''({\bf 5})$,$\bar{\Phi}''({\bf 5^*})^{\dagger}$  \\
\hline
  Charges & 0 & $-1$ & $-3$ & 2 & 2 
 \end{tabular}
\caption{Fiveness charge for each hypermultiplet.}
\label{tab:fiveness}
\end{table}

\end{document}